\documentclass[prd,amssymb,amsmath,floatfix]{revtex4}
\pdfoutput=1 
\usepackage{graphicx}
\usepackage{natbib}

\begin{document}
\title{From Cosmos to Intelligent Life: The Four Ages of Astrobiology}

\author{Marcelo Gleiser\footnote{mgleiser@dartmouth.edu. Opening plenary talk delivered at the S\~ao Paulo Advanced School of Astrobiology, S\~ao Paulo, December 2011.}}
\affiliation{Department of Physics and Astronomy, Dartmouth College
Hanover, NH 03755, USA}

\begin{abstract}
The history of life on Earth and in other potential life-bearing planetary platforms is deeply linked to the history of the universe. Since life as we know it relies on chemical elements forged in dying heavy stars, the universe needs to be old enough for stars to form and evolve. Current cosmological theory indicates that the universe is 13.7$\pm 0.13$ billion years old and that the first stars formed hundreds of millions of years after the big bang. At least some stars formed with stable planetary systems wherein a set of biochemical reactions leading to life could have taken place. In this lecture, I argue that we can divide cosmological history into four ages, from the big bang to intelligent life. The {\it physical age} describes the origin of the universe, of matter, of cosmic nucleosynthesis, as well as the formation of the first stars and galaxies. The {\it chemical age} begun when heavy stars provided the raw ingredients for life through stellar nucleosynthesis and describes how heavier chemical elements collected in nascent planets and moons to give rise to prebiotic biomolecules. The {\it biological age} describes the origin of early life, its evolution through Darwinian natural selection, and the emergence of complex multicellular life forms. Finally, the {\it cognitive age} describes how complex life evolved into intelligent life capable of self-awareness and of developing technology through the directed manipulation of energy and materials. We conclude discussing whether we are the rule or the exception.
\end{abstract}

\maketitle

\section{Introduction}

\noindent
During the past decades a growing influx of data has fed a twin revolution in the cosmic sciences: on the cosmological scale, the Cosmic Background Explorer (COBE) \cite{COBE} and the Wilkinson Microwave Anisotropy Probe (WMAP) \cite{WMAP} have mapped the properties of the cosmic microwave background to remarkable accuracy, allowing cosmologists to answer age-old questions related to the universe: how old it is, what is its geometry, what is its material composition, and when did the first stars appear. Coupled to large-scale telescopic surveys that mapped the distribution of galaxies in the universe such as the Sloan Digital Sky Survey (SDSS) \cite{SDSS} and the 2dF Galaxy Redshift Survey (2dF) \cite{2dF}, as well as data from the Hubble Space Telescope Key Project \cite{Freedman} and many other completed and on-going surveys, there is widespread support for the so-called cosmic concordance cosmological model, or $\Lambda$CDM model. According to this model, the universe is 13.7 billion years old and consists of 4\% baryonic mater, 23\% dark matter, and 73\% dark energy. The Hubble constant characterizing the cosmic expansion is 71 km/s/Mpc and the density of the universe is very close to the critical value for a flat geometry.

In addition, within our galactic neighborhood a combination of different observational techniques has led to the discovery of hundreds of exoplanets. As of 4 February 2012, a total of 758 such planets have been found, and a recent study estimates that each star of the 100 billion or so in the Milky Way hosts on average 1.6 planets \cite{Cassan}. This being the case, we should expect hundreds of billions of planets (to say nothing of their possible moons) in our galaxy alone. Although most discoveries thus far were made with ground-based telescopes using Doppler shift, transit, and gravitational microlensing techniques, two satellites using transit techniques, Corot (launched December 2006) \cite{COROT} and Kepler (launched March 2009) \cite{KEPLER}, have quickly added to the growing numbers. Kepler, in particular, was designed to search for Earth-like planets in or near the habitable zone of our galaxy. As of December 2011 the Kepler team had identified 2,326 candidates, with 207 of these having similar masses to Earth. That same month, two Earth-sized candidates were confirmed orbiting a star only slightly smaller than the Sun (91\% of the Sun's mass) in the constellation of Lyra, some 950 light-years from Earth. Unfortunately, both planets have orbits closer to their star than that of Mercury to the Sun and thus surface temperatures much higher than what living organisms can bear. Still, the Kepler team estimates that 5.4\% of all stars host Earth-sized candidates. 

Taken together, the cosmological and exoplanetary data indicate that there are plenty of potentially life-bearing platforms within our galaxy. Based on the fact that the same laws of physics and chemistry apply throughout the universe, the same conclusion can be extended to the hundreds of billions of other galaxies within our cosmological horizon. We can thus organize the history of life in the universe in terms of the steps needed for matter to have sequentially self-organized into more and more complex structures, from the first atomic nuclei into stars and planets (Physical Age), from heavier chemical elements into biomolecules (Chemical Age), into living creatures of growing complexity (Biological Age), and, finally, into thinking assemblies of biomolecules and possibly beyond (Cognitive Age). This way, it is clear that the history of life in the universe--a central concern of astrobiology--begins with the origin of the universe itself. In what follows, we will briefly present some of the key aspects of each of these ages.

\section{The Physical Age}

\noindent
If one considers the possibility of a multiverse seriously, and there are good reasons to do so both from inflationary cosmology \cite{Linde1, Vilenkin1} and from string theory and the related notion of a landscape of an enormous number of possible solutions \cite{Susskind,Weinberg1}, our universe is one of many (infinitely many?) cosmoids that constantly bubble forth from a timeless realm. Multiverse or not, our universe does need to satisfy a few properties in order to support the increasing complexification of matter from the most elementary particles to atomic nuclei and light atoms, and from these to the first stars and star-forming regions. This era of cosmic emergence and of the formation of various bound states of matter defines the Physical Age of astrobiology, which began with the big bang and is ongoing today. (In fact, the natural processes defining each age, once started, have remained active until today and will continue for the foreseable future. If they will remain active into the far distant future depends on how dark energy will influence the ultimate fate of the universe \cite{BigRip}.) 

Current cosmological theory, with support from observations, have established that our universe needs a vacuum energy density (aka dark energy), matter-antimatter asymmetry, dark matter density, and primordial perturbations for the formation of large-scale structure, probably generated during an early period of rapid expansion known as inflation. To these, one adds the couplings of the four interactions (gravitational, electromagnetic, and strong and weak nuclear forces) and the masses of quarks and leptons so that hadrons and then light nuclei can form after electroweak symmetry breaking. The early universe is only consistent with a small initial entropy $S_i$ \cite{Carroll}, since large-entropy initial states will not be conducive of structure formation of any kind: for complexity to emerge there must be enough free energy \cite{Entropy}.

The Physical Age thus includes all of the early history of the universe, from the big bang to electroweak symmetry breaking to primordial nucleosynthesis to recombination at roughly 400,000 years, when the first hydrogen atoms were formed and the photons' mean-free path became comparable to the causal horizon: the cosmic microwave background was born. Across the expanding cosmic volume there were overdense regions, where fluctuations from primordial inflation had gathered enough dark matter to create long overdense sheets and filaments and, in some regions, deeper gravitational potential wells for baryonic matter to fall in and thus form large hydrogen-rich clouds. After about 200 million years, these clouds had contracted gravitationally and accreted enough material to ignite nuclear fusion within their cores: the first supermassive stars were born \cite{Abel}. Galaxies soon followed or co-evolved with the first supermassive stars. Recent observations have found galaxies when the universe was just 480 millions years old \cite{FirstGalaxy}. 

Fast forward a few hundreds of millions of years, and galactic evolution and mergers were happening in earnest. Life-supporting galaxies, however, must satisfy a few constraints in their morphology and stellar ages and types.  In order to retain heavy elements, galaxies must have masses above a certain value. Numerical results suggest that galaxies with $M\geq 10^9M_{\odot}$ are able to retain $\gtrsim 30\%$ of their heavy elements \cite{LowFerrara}. Merging processes to form galactic disks such as that of the Milky Way (which has a mass $M_{\rm MW}\sim 10^{12}M_{\odot}$), take time to complete and are unlikely for $z > 1$ \cite{AbrahamBergh}. So, life is favored in large-mass galaxies, and we could take $M_{\rm MW}$ as a fiducial value. Stars can be constrained by type, since continually habitable zones (i.e., surface liquid water for extended periods) exist only around stars in the spectral classes between F5 to mid-K \cite{Kasting1,Livio}, which amount to about $\sim 20\%$ of main sequence stars. Of these, we must include only the fraction bearing planetary systems. Current searches for exoplanets indicate that this fraction can be substantial. Perhaps surprisingly, even binary star systems may contribute. I will leave the notion of a galactic habitable zone aside.

Once there are high metallicity stars and planetary disks sprinkled with the chemical elements needed for life the Chemical Age can begin.

\section{The Chemical Age}

\noindent
Since a rich variety of inorganic and organic molecules has been found in the interstellar medium through spectroscopy \cite{MoleculesList}, we are certain that chemistry need not be constrained to planetary platforms or their atmospheres. One of the key open questions related to the origin of life is precisely whether the first ingredients for life may have been delivered to early Earth, either through meteoritic impact \cite{Cronin,Pizzarello} or simply by constant deposition \cite{ChybaSagan}, or whether they were synthesized here \cite{Miller}: amino acids have been found in carbonaceous chondrites and have been synthesized in the laboratory from simple chemical blocks and conditions simulating Earth's primeval atmosphere. Indeed, it is likely that life on Earth took advantage of both local synthesis and delivery. 

Of the planets in the habitable zone, only a fraction will have the right preconditions for life: liquid water and the elements C, O, H, N, and the less abundant but no less needed P, S, Fe, Ca, Na, Cl, etc. Apart from water, other simple molecules are also supposed to be present, although the specifics may differ (CH$_4$, CO$_2$, NH$_3$...). As indicated by Miller-type experiments \cite{Miller}, to produce amino acids {\it in situ} there is also a need for a reducing atmosphere. Planetary platforms with volcanism have a clear advantage \cite{Bada}. Taking the age of the Earth to be 4.54 billion years \cite{EarthAge} and the Late Heavy Bombardment to have taken place between 4.1 and 3.8 billion years ago \cite{Bombardment}, there is little that can be said with certainty about what course prebiotic chemistry may have taken during these first 740 million years of Earth's history. For example, Davies and Lineweaver conjectured that there may have been several early life experiments, which were reset by intense environmental disturbances \cite{Davies}. Given that such considerations somewhat blur the line between the chemical and biological ages, we adopt a working hypothesis that the chemical age is characterized by the prebiotic chemical processes that led to the first successful life experiment, irrespective as to when it happened. Clearly, by 3.5 to 3.4 billion years ago, when there is convincing evidence for life being present on Earth in the form of single-celled prokaryotes, the biological age had begun in earnest. Thus, the fundamental open question that lies at the boundary of the two ages is precisely the transition form nonliving to living matter, or abiogenesis. 

If we adopt the working (and necessarily oversimplistic) definition of life as a self-sustaining network of chemical reactions capable of exchanging energy with the environment and of Darwinian reproduction, prebiotic chemistry addresses the emergence of such a network of reactions. In a general sense, chemistry describes matter's urge to bond in an attempt to decrease asymmetries in atomic and molecular electric charge distributions. Life is a very complex manifestation of this urge, an imbalance that recreates itself \cite{Tear}:  it is not matter, but a process that happens to matter. Although the uniformity of life on Earth suggests that all extant organisms descended from a last universal common ancestor (LUCA), we know little of the abiotic ingredients and prebiotic chemistries present on the primitive Earth from which the LUCA evolved \cite{O}. Potential mechanisms range from ``metabolism-first'' models, such as the iron-sulfide world hypothesis of W\"achtersh\"auser \cite{Wach} and ``membrane--first'' lipid-world scenarios investigated by Deamer and coworkers \cite{MD,MBD}, to the ``peptide-first'' models proposed by Fox \cite{Fox,Fox2} and others \cite{Fishkis,GW}, and the popular ``genetics-first'' hypotheses such as in the RNA \cite{Gilbert} and pre-RNA \cite{Orgel2000} world scenarios. These mechanisms could generally be grouped into the ``metabolism-first,'' inspired by Oparin's seminal work,  and ``genes-first'' schools. Coupled to the question of abiogenesis is the origin of homochirality, or why biomolecules display a near perfect spatial asymmetry \cite{Chiral,Gleiser}.

In his 1924 book {\it The Origin of Life} \cite{Oparin}, Oparin noted that drops of oily liquids don't generally mix well with water, forming small bubble-like droplets instead. These fatty droplets, according to Oparin, would have made a nice protective environment allowing molecules accidentally trapped in their interior to react with each other with reduced external interference \cite{MD,MBD}. Occasionally, certain reactions would produce more chemicals and grow in complexity. At a critical threshold, the molecules would be able to produce more copies of themselves in a self-sustaining (``autocatalytic'') reaction network: the little fatty bags would become the first protocells. As opposed to reproduction in a more organized genetic framework, reproduction here would initially happen at random, as turbulent external conditions would force some droplets to split. (Shaking salad dressing shows this.) In rare cases, the daughters would contain the right chemicals to also maintain self-sustaining reactions and a population of somewhat similar protocells would start to develop. Doron Lancet and collaborators at the Weitzman Institute have developed computer simulations of such Òlipid worldÓ scenarios, showing that when a parent cell can produce more than one self-catalyzing daughter a chain reaction may occur, leading to a kind of primitive life \cite{Lancet}. Genetics would develop later, as the reproductive process perfects itself through countless ``generations,'' led by the invisible hand of some prebiotic version of natural selection. We should expect that protocells containing molecules that reproduced more efficiently and that could better extract and metabolize energy from the outside environment had an advantage over others and slowly came to dominate the population \cite{Tear}.

The countering position is that genetics came first: duplication preceded metabolism. The most popular idea within this view is the ``RNA world'' hypothesis \cite{Gilbert}: of the two genetic information carriers, DNA and RNA, RNA is the one with the ability to jump-start its own duplication process. Unlike DNA, it can function as an enzyme, so it is able to catalyze its own polymerization (that is, the chaining of smaller pieces into longer molecules like pearls in a necklace) and duplication. If we assume, quite reasonably, that life started simple, a self-sufficient replicator is one way to go. As Tom Fenchel remarked in {\it Origin and Early Evolution of Life} \cite{Fenchel}, the real advantage of the RNA-first scenario is that it allows for extensive laboratory-based studies. Many remarkable experiments, such as those by Manfred Eigen and Leslie Orgel \cite{Eigen} and, more recently, by Gerald Joyce's group \cite{Joyce2} at the Scripps Research Institute in San Diego, California, have clarified the relationship between genetics and natural selection at the molecular level through direct RNA and DNA manipulation, illuminating the connection between chemistry and biology. However, from the point of view of life's origins, it should be clear that for RNA to be present in early Earth a lot of complex chemical syntheses had already taken place. 

Quite possibly, as Dyson suggested in {\it Origins of Life} \cite{Dyson}, both scenarios worked to generate the first ``thing'' we could call living; at some point, protocells with primitive metabolism and simple lipid boundaries--the cellular hardware--were invaded by or accidentally absorbed the precursors of genetic replication--the cellular software--as parasites invade a host.  After eons of trial and error, a symbiotic fusion of the two eventually developed, creating a cell with optimized replication capability. In any case, this discussion illuminates the point made learner that the boundaries between the chemical and biological ages are quite blurry. We may thus assume that, at some point between 3.8 and 3.5 billion years ago the LUCA appeared and the biological age started in earnest. Of course, it may have started eons before in another planetary platform in this or other galaxy. In any case, the transition from nonliving to living matter would have happen there as well--unless life was delivered ready-made from space. However, even if one accepts the panspermia hypothesis \cite{Panspermia,Spores} (this author, in particular, finds it quite far-fetched), abiogenesis had to happen at least once in the universe.

\section{The Biological Age}

\noindent
Once life begins--or even before, at the biomolecular, prebiotic level--Darwinian natural selection is at work. The simplest autonomous living entity is a cell. There is, of course, an enormous and ill-understood jump in complexity from coacervates with some kind of duplicating software to the simplest cells known to us, prokaryotes. Blue-green algae and many bacteria are prokaryotes, primitive cells where DNA is bundled into a coil without a membrane separating it from the rest of the cell. In eukaryotes, the more sophisticated cells like the ones in our body, an isolated nucleus houses the genetic material. As we look into the history of life on Earth, we discover that single-celled organisms were by far its most enduring inhabitants. The numbers are remarkable: from around 3.5 billion years ago, life remained unicellular until about 1 billion years ago. That is, for roughly 2.5 billion years, life on Earth consisted only of single-celled organisms, albeit some organized in colonies. Eukaryotes appeared close to the end of this period, thanks to collective efforts of the photosynthetic blue-green algae, when oxygen became more abundant in the atmosphere.

This fact should give us pause. To study life's origin we can forget about multi-cellular organisms. The stars are the prokaryotes. The crucial transition from single-celled to multi-celled organisms, from our amoeba-like ancestors to sponges, happened for a number of improbable factors: most importantly, the increase in atmospheric oxygen between 2.7 and 2.2 billion years ago. A consequence of this increase is the parallel production of ozone due to the action of UV sunlight on oxygen. This ozone created a protective layer between organisms and the same nasty UV radiation from the Sun, allowing more complex life forms to evolve. We wouldn't be here without it. When we consider the possibility of life elsewhere in the cosmos these factors (and many more) are crucial. The key point to keep in mind is that the history of life on any planet (or moon) is deeply related to the planet's geological history. Put it a different way, had we rewound the clock on Earth's history and changed a few major events in its geological history, life would have taken a very different turn. Mutations occur at random and have no ``hidden agenda'' toward increased complexity: all life cares about is adaptability. If life forms are well-adapted to the environment, that is, if there is little or no environmental pressure, mutations will hardly be beneficial. (They mostly aren't anyway.) Thus, different environmental pressure will lead to different demands on adaptability and hence on the phenotype of successful life forms \cite{Lunine}.

From the perspective of astrobiology, both the study of life on Earth (from its origins to present-day extremophiles \cite{Baross}) and the possibility of detecting the signature of life in exoplanets \cite{Kaltenegger} have been the focus of much activity and the basis for the planning of future exploratory missions and telescopic design \cite{Telescopes}. See, e.g., the books by Sullivan and Baross \cite{Baross}, and by Lunine \cite{Lunine} for some of the key questions being investigated at the moment. Here, I end this section on the biological era by listing some of the key steps toward increasing complexification that happened on Earth (dates are approximate):

$\bullet$ From prokaryotic to eukariotic cells (probably through endosymbiosis \cite{Margulis}) (2 billion years ago)

$\bullet$ From eukaryotic cells to colonies and multicellular organisms (1 billion years ago)

$\bullet$ From multicellular organisms to simple animals, arthropods and complex animals (600-550 million years ago)

$\bullet$ Diversification of life into fish, amphibians, land plants, insects, reptiles, mammals, birds, flowers to dinosaur extinction (500-65 million years ago)

$\bullet$ Appearance of the genus {\it Homo} (2.5 million years ago)

$\bullet$ Appearance of anatomically modern humans (200,000 years ago)\\

On Earth at least, the advent of complex multicellular life led, after about 500 million years of evolution, to the appearance of beings capable of self-awareness and of manipulating their environment to create tools to enhance their quality of life. Although it can be argued that bonobos, chimpanzees, dolphins and other mammals display a high level of self-awareness, emotional depth, and crude tool manipulation, I am defining the dawn of the cognitive age to coincide with the dawn of the first modern humans, the only species we know of capable of creating complex technology based on the assembly of different materials {\it and} of creating art, that is, the only species that has both functionality and aesthetic considerations during the act of creation extending beyond mere survival needs.

\section{The Cognitive Age}

\noindent
The prevalent question of our age, at least when it comes to our place in the universe, is whether we are the rule or the exception. Given the plurality of worlds and the abundance of organic chemicals in the interstellar and circumstellar medium (for example, the recently-found polycyclic aromatic hydrocarbons \cite{PAH}), added to the remarkable resilience of terrestrial extremophiles, it is hard to support the idea that life has only found its way on our planet. On the other hand, when thinking about life in the universe one must distinguish between simple, one-celled life and complex, multicellular life. And even here, we must be careful to distinguish between multicellular and intelligent life. Taking Earth as our only illustration thus far, life was single-celled for about 3 billion out of 3.5 billion years. Intelligence, very broadly defined as the ability to fashion tools for a definite purpose, only for the past million years or so with {\it Homo Abilis}, although more complex tool making, the interest and ability to bury the dead, and the advent of art--three characteristics of the cognitive age-- probably came only much more recently. Here, we must add an important remark: given life's dependence on the environment--{\it the history of life in a planet mirrors the planet's life history}--life as we know it on Earth cannot be replicated. However, we must still wonder whether intelligence is a reproducible feature of life, that is, whether life elsewhere can be intelligent, capable of art and technology. And, if so, whether it is widespread in the cosmos.

In their courageous {\it Rare Earth: Why Complex Life Is Uncommon in the Universe}, Peter Ward and Donald Brownlee argued very convincingly that life may not be uncommon in the Universe but it likely exists elsewhere only in its simplest form: alien Earth-like planets may support alien microorganisms but not much more than that \cite{Ward}. Complex, multicellular life relies on too many planetary factors--even after clearing all the chemical roadblocks--to be common. (For example, a large moon to stabilize the planetary axis tilt and hence the seasons, a magnetic field to shield off radiation, plate tectonics to remix surface and ocean chemistry that helps regulate CO$_2$ levels, etc.)

Since it is difficult to imagine how intelligence--here or anywhere else--could have emerged without millions of years of evolving multicellular creatures, the discovery of multicellular aliens would be a great boost to the possibility that there are other smart creatures out there. Even so, it is important to keep in mind that human intelligence appeared as a by-product of random cosmic and evolutionary accidents: intelligence is not the end-goal of evolution, as one hundred and fifty million years of dinosaurs demonstrate \cite{Tear}.

 If we take the possibility of alien intelligent life seriously, we must ponder a few questions. The most obvious, as Enrico Fermi noted in 1950, is ``Where is everybody?'' Our galaxy is about 13.2 billion years old, almost three times as old as the Sun. If we imagine that life evolved in another stellar system even as little as a few million years earlier than it did here, and that it reached a stage in its evolution where complex creatures became intelligent, then it follows that some of these aliens would have had plenty of time to reach amazingly advanced levels of technological sophistication. Considering what we have achieved with only four hundred years of modern science, their technology would be like magic to us. If, like humans, they suffer from wanderlust, they would have had the means and plenty of time to explore the galaxy many times over. (Traveling at 0.1c, it would take one million years to cross the galaxy.) And yet, the evidence at hand indicates that they haven't colonized the galaxy or visited us here on Earth. So, where is everybody? This issue is sometimes called Fermi's Paradox. Fifty possible resolutions, some amusing and others quite serious, can be found in Ref.  \cite{FermiParadox}. 
 
It is entirely possible that other intelligent life forms exist and have existed before us. The Cognitive Age started when the first signs of intelligence appeared in a corner of our causally-connected universe. If they exist within our galactic neighborhood, say, within a few hundred light-years away, we stand a chance to ``listen in'' on their radio transitions, if any. This is the goal of the SETI program, which has been in operation for over half a century \cite{SETI}. A preferred frequency window is the Microwave Window, between 1 and 10 GHz, a range where signals travel quite unimpeded across gas and dust. Within this window, many searches focus on the window from 1420 MHz (hydrogen 21 cm line) and 1720 MHz (the higher of the four hydroxyl molecule frequencies). Water forms when hydrogen combines with hydroxyl. Since water is an essential ingredient for life, this range is known as the ``Water Hole''. Of course, there is a strong assumption here, that aliens would want to send signals within this specific frequency range to communicate with other technological civilizations. Although the odds that a successful discovery will be made this way are small, the high pay off certainly justifies the effort. Quoting Jill Tarter, current director of the SETI InstituteÕs Center for SETI Research, "if you don't look you won't find." Currently, there are several ways in which SETI is searching for evidence of intelligence, including planetary-scale engineering projects and other possible signs of purposeful redesigning at astronomical scales \cite{DaviesAliens}. Of course, we can always speculate that a sign of high intelligence is to have developed ways to mask one's presence so as not to be bothered by lesser species.

As of this writing, we have no evidence of life elsewhere. It is widely expected that this will change within the next few decades, through spectroscopic evidence of Earth-like exoplanetary atmospheres, where ozone or even chlorophyl and other tell-tale signatures of life may be detected \cite{Kaltenegger}. More direct searches in the subglacial oceans of Europa, or even in subterranean Mars, may find some evidence of past or present simple life forms. In all likelihood, intelligent life will be a much rarer find. Given the vast distances involved in interstellar travel, we may not find the answer in the foreseeable future. However, as Carl Sagan was fond of saying, ``absence of evidence is not evidence of absence.'' Search we must, if only as a directive of our own intelligence.


\section*{Acknowledgements}

MG is supported in part by a National Science Foundation grant PHY-1068027. 

\bibliographystyle{}

\end{document}